\documentclass[12pt]{iopart}
\usepackage{graphicx}
\begin{document}

  \newcommand {\nc} {\newcommand}
  \nc {\beq} {\begin{eqnarray}}
  \nc {\eeq} {\nonumber \end{eqnarray}}
  \nc {\eeqn}[1] {\label {#1} \end{eqnarray}}
  \nc {\eol} {\nonumber \\}
  \nc {\eoln}[1] {\label {#1} \\}
  \nc {\ve} [1] {\mbox{\boldmath $#1$}}
  \nc {\ves} [1] {\mbox{\boldmath ${\scriptstyle #1}$}}
  \nc {\mrm} [1] {\mathrm{#1}}
  \nc {\half} {\mbox{$\frac{1}{2}$}}
  \nc {\thal} {\mbox{$\frac{3}{2}$}}
  \nc {\fial} {\mbox{$\frac{5}{2}$}}
  \nc {\la} {\mbox{$\langle$}}
  \nc {\ra} {\mbox{$\rangle$}}
  \nc {\eq} [1] {(\ref{#1})}
  \nc {\Eq} [1] {Eq.~(\ref{#1})}
  \nc {\Ref} [1] {Ref.~\cite{#1}}
  \nc {\Refc} [2] {Refs.~\cite[#1]{#2}}
  \nc {\Sec} [1] {Sec.~\ref{#1}}
  \nc {\chap} [1] {Chapter~\ref{#1}}
  \nc {\anx} [1] {Appendix~\ref{#1}}
  \nc {\tbl} [1] {Table~\ref{#1}}
  \nc {\Fig} [1] {Fig.~\ref{#1}}
  \nc {\ex} [1] {$^{#1}$}
  \nc {\Sch} {Schr\"odinger }
  \nc {\flim} [2] {\mathop{\longrightarrow}\limits_{{#1}\rightarrow{#2}}}

\title[Extracting nuclear structure information from the breakup of exotic nuclei]{Open issues in extracting nuclear structure information from the breakup of exotic nuclei}

\author{P.~Capel}

\address{Physique Nucl\'eaire et Physique Quantique (CP 229)\\
Universit\'e Libre de Bruxelles (ULB)\\
50 av. F.~D.~Roosevelt, B-1050 Brussels, Belgium}
\ead{pierre.capel@ulb.ac.be}
\begin{abstract}
The open issues in the development of models for the breakup of exotic nuclei and the link with the extraction of structure information from experimental data are reviewed.
The question of the improvement of the description of exotic nuclei within reaction models is approached in the perspective of previous analyses of the sensitivity of these models to that description.
Future developments of reaction models are suggested, such as the inclusion of various channels within one model.
The search for new reaction observables that can emphasise more details of exotic nuclear structure is also proposed.
\end{abstract}

\pacs{24.10.-i, 25.60.-t, 21.10.Gv, 25.60.Gc, 25.60.Bx}

\section{Introduction}
The development of radioactive-ion beams (RIB) in the mid-eighties has enabled the exploration of the nuclear landscape away from stability.
This technical breakthrough has revealed the existence of unexpected exotic nuclear structures near the driplines such as halo nuclei \cite{Tan85r,Tan85b} or shell inversions \cite{WBB90}.
Halo nuclei are neutron-rich nuclei that exhibit a significantly larger matter radius than their isobars.
This exceptional size of these nuclei is now understood as being due to their low binding energy for one or two neutrons \cite{Tan96, Jon04}.
Thanks to this lose binding, the valence neutrons can tunnel far away from the other nucleons and exhibit a high probability of presence in the classically forbidden region \cite{HJ87}.
They thus form a sort of \emph{halo} around the core, which exhibits the same characteristics (size, density\ldots) as stable nuclei.
Proton haloes are also possible, though less probable due to the presence of a Coulomb barrier, which hinders the formation of a long tail in the nuclear density.

Being located away from stability halo nuclei cannot be studied through usual spectroscopic techniques and one must rely on indirect methods such as nuclear reactions to infer information about their exotic structure.
The best known reactions used to probe the nuclear structure far from stability are elastic scattering \cite{Dip10}, breakup \cite{Fuk04}, knockout \cite{Sau00} and transfer \cite{Jon10}.
The breakup reaction is particularly well suited to study loosely-bound systems, such as halo nuclei.
In that reaction, the nucleus under study is sent on a target and events are studied, in which the projectile breaks up into its constituents, e.g.\ the halo neutrons and the core, hence revealing its structure.
Because all final fragments are detected in coincidence, one often speaks about \emph{elastic} or \emph{diffractive breakup}.

To infer valuable nuclear-structure information from reaction data, an accurate reaction model coupled to a realistic description of the projectile is needed.
Various such models have been developed: the coupled-channel technique with a discretised continuum (CDCC) \cite{Kam86,Yah86,TNT01}, the time-dependent model (TD) \cite{KYS94,EBB95,TW99,CBM03c,LSC99}, and models based upon the eikonal approximation \cite{Glauber}, such as the eikonal-CDCC (E-CDCC) \cite{OYI03}, the dynamical eikonal approximation (DEA) \cite{BCG05} or the Coulomb-corrected eikonal model (CCE) \cite{MBB03,AS04,CBS08} (see \Ref{BC12} for a recent review).
The goal of this contribution is to present the various issues that need to be addressed in order to better extract structure information from breakup reactions.
After a brief reminder of the theoretical framework, the major issue of the improvement of the projectile description in existing reaction models is presented in \Sec{structure}.
Then, ideas to extend the validity range and/or reduce the computation cost of existing models are listed in \Sec{reaction}.
In \Sec{observable}, the development of new reaction observables that are more sensitive to the projectile structure is suggested.
The example of the recent ratio method \cite{CJN11} is discussed.
An outlook is proposed in \Sec{conclusion}.

\section{Theoretical framework}\label{theory}
\begin{figure}
\center
\includegraphics[width=4cm]{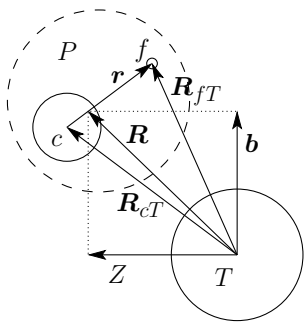}
\caption{Jacobi set of coordinates used to describe the collision of a two-body projectile $P$ on a target $T$.
The components of the $P$-$T$ relative coordinate $\ve{R}$ longitudinal $Z$ and transverse $\ve{b}$ to the beam axis are displayed.}\label{f1}
\end{figure}

The theoretical description of reactions involving loosely-bound nuclei is usually expressed in the following few-body framework.
The projectile $P$, i.e.\ the nucleus under study, is described as a two- or three-body system: an inert core $c$ to which one or two valence particles, denoted as the fragment(s) $f$, are loosely bound.
For a two-body projectile, e.g.\ a one-neutron halo nucleus, the internal structure is described by the Hamiltonian
\beq
H_0=-\frac{\hbar^2}{2\mu}\Delta_{\ve{r}}+V_{cf}(r),
\eeqn{e1}
where $\mu$ is the $c$-$f$ reduced mass and $\ve{r}$ is the relative coordinate of the fragment to the core (see \Fig{f1}).
The parameters of the phenomenological potential $V_{cf}$ are adjusted to physical data: the binding energy of the system, the energy of excited states, which may be bound or correspond to resonances in the $c$-$f$ continuum, the spin and parity of these states etc.
The eigenstates $\Phi$ of Hamiltonian $H_0$ describe the $c$-$f$ relative motion.
The negative-energy states correspond to the bound states of the system, while the positive-energy states simulate the $c$-$f$ continuum.

In most reaction models, the target is seen as a structureless particle, which interacts with the projectile constituents through optical potentials chosen in the literature: $U_{cT}$ and $U_{fT}$.
Within this framework, studying the $P$-$T$ collision reduces to solving the three-body \Sch equation
\beq
\left[-\frac{\hbar^2}{2\mu_{PT}}\Delta_{\ve{R}}+H_0+U_{cT}(\ve{R}_{cT})+U_{fT}(\ve{R}_{fT})\right]\Psi(\ve{R},\ve{r})=E\Psi(\ve{R},\ve{r}),
\eeqn{e2}
with $\mu_{PT}$ the $P$-$T$ reduced mass and $E$ the total energy in the $P$-$T$ centre-of-mass restframe.
\Eq{e2} must be solved with the boundary condition that the projectile is impinging on the target in its ground state $\Phi_0$:
\beq
\Psi(\ve{R},\ve{r})\flim{R}{\infty}e^{iK_0Z}\Phi_0(\ve{r})+\mbox{outgoing waves},
\eeqn{e3}
where $K_0$ is the wave number of the initial $P$-$T$ relative motion.
It is related to the total energy $E$ and the energy $\varepsilon_0$ of the initial bound state following $E=\hbar^2K_0^2/2\mu_{PT}+\varepsilon_0$.
The outgoing waves describe the scattering of the projectile by the target in any of the eigenstates $\Phi$ of the Hamiltonian $H_0$ \eq{e1}.
This includes elastic scattering if $\Phi=\Phi_0$, inelastic scattering if $\Phi$ is another bound state of $H_0$, and breakup if $\Phi$ is a positive-energy state of $H_0$.

Although this equation could in principle be exactly solved within the Faddeev technique, only a few cases can actually be handled by this method.
Due to a divergence problem induced by the $P$-$T$ Coulomb interaction, only (very) light targets can be considered up to now \cite{DFS05,DFS05b}.
In most cases, approximations must be performed to solve \Eq{e2}.
Various models have been developed to do so.
Below the models mostly used for the analysis of experimental data are briefly presented.
The interested reader is referred to \Ref{BC12} for more details.

In the continuum discretised coupled channel model (CDCC), the wave function $\Psi$ is expanded over the eigenstates $\Phi$ of the projectile Hamiltonian $H_0$.
This method leads to a set of coupled equations \cite{Yah86}.
The major approximation of CDCC lies in the truncation of the set of $\Phi$ used to compute the wave function $\Psi$ (see Refs.~\cite{Kaw86,AYK89,AKY96} for more details on the theoretical foundations of CDCC).
Since no assumption is made on the projectile-target relative motion CDCC is valid on the whole beam-energy range.
However, to reach convergence, CDCC requires a rather large model space, which can be computationally challenging, especially at large energy \cite{CEN12,Tho14}.
This hinders the extension of the CDCC framework to descriptions of the projectile beyond the simple two-body model mentioned above (see \Sec{structure}).
Since the channel of interest corresponds to the dissociation of the halo from the core, the projectile continuum must be included in this expansion in a tractable way.
One then uses a set of square-integrable states that simulate the continuum.
Usually, these states are obtained by the \emph{binning} (or average) technique, in which exact continuum states are averaged over small intervals of continuum energies, or \emph{bins} \cite{Yah86}.
Pseudostates obtained by diagonalising $H_0$ within a finite basis of square-integrable functions such as Gaussians \cite{Mat03}, by transformation of a harmonic-oscillator basis (THO) \cite{PBM01} or within the $R$-matrix technique \cite{DBD10} can also be used.

Other models make assumptions on the projectile-target relative motion.
This greatly simplifies the calculations but limits the models in energy range and/or breakup observables that can be reliably described.
The time-dependent model (TD) relies on the semiclassical approximation in which the projectile-target relative motion is described by a classical trajectory \cite{AW75}.
Along that trajectory, the projectile feels a time-dependent potential that simulates its interaction with the target.
This approximation leads to the resolution of a time-dependent \Sch equation.
Various algorithms have been developed to solve numerically this time-dependent equation for two-body projectiles \cite{KYS94,EBB95,TW99,CBM03c,LSC99}.
Thanks to this simplification in the treatment of the $P$-$T$ relative motion, the computational effort of the TD technique is much lower than that of CDCC.
However, relying on the semiclassical approximation, quantal effects, such as interferences observed in angular distributions, cannot be reproduced \cite{CEN12}.

At sufficiently high energy, the eikonal approximation can be used to describe nuclear reactions \cite{Glauber}.
In that approximation the $P$-$T$ relative motion is supposed not to vary much from the incoming plane wave \eq{e3}.
The idea is thus to factorise that plane wave out of the wave function $\Psi$ and consider what remains as smoothly varying with $\ve{R}$, hence simplifying the \Sch equation \eq{e2}.
Both the eikonal CDCC (E-CDCC) \cite{OYI03} and the dynamical eikonal approximation (DEA) \cite{BCG05} make use of that idea.
In the former, the eikonal equation is solved by expanding the wave function $\Psi$ on the projectile eigenstates $\Phi$, exactly as in CDCC.
E-CDCC is nevertheless much more computationally tractable thanks to the underlying eikonal approximation.
In DEA, the wave function is expanded over a three-dimensional mesh, which enables a faster convergence than the continuum discretization \cite{FOC14}.
Unlike the TD model, the eikonal approximation is fully quantal and hence can reproduce the interferences in angular distributions \cite{GBC06}.
However being a high-energy approximation, it cannot be used to analyse breakup observables at too low energy, e.g.\ from ISOL experiments.
A recent comparison between CDCC, TD and DEA details these differences \cite{CEN12}.

The usual eikonal approximation assumes a subsequent adiabatic or sudden approximation to the E-CDCC and DEA.
This corresponds as seeing the projectile to be frozen during its interaction with the target.
While valid for the short-range nuclear interaction, the sudden approximation is incompatible with the (infinite-range) Coulomb force.
The usual eikonal approximation hence diverges when applied to the Coulomb breakup of loosely-bound nuclei \cite{GBC06}.
To circumvent this problem, Margueron, Bonaccorso and Brink have introduced a first-order correction to the treatment of the Coulomb interaction within the usual eikonal model \cite{MBB03}.
This Coulomb-corrected eikonal model (CCE), studied in \Ref{AS04}, has been confronted to the DEA in \Ref{CBS08}.
The ability of the CCE to reproduce both Coulomb- and nuclear-dominated reactions coupled to its computational simplicity make it ideal to extend reaction models to three-body projectiles \cite{BCD09,PDB12} (see \Sec{4body}).

\section{Projectile description}\label{structure}

\subsection{Microscopic description of the projectile}\label{micro}

The major issue that should be addressed in future developments in breakup modelling is that of the projectile description.
As mentioned in \Sec{theory}, most current reaction models treat the projectile as a simple two-body system: a valence particle loosely-bound to an inert core.
The analyses of reactions which aim at disentangling the various states in which the core can be in the projectile ground state, or which probe more complex projectile structures, such as two-nucleon halo nuclei, require a finer description of the projectile than this simple single-particle structure.

The ultimate goal would be to combine a microscopic description of the projectile within the CDCC reaction model.
In this way, one would be able to study at all beam energies the breakup of any kind of nucleus, naturally including all possible structure channels.
Although some efforts have already been made in this direction \cite{DH13}, they remain limited to the elastic-scattering channel. 
Moreover it is not clear that such a model is actually needed.
Indeed, previous studies have shown that breakup, both Coulomb and nuclear dominated, is a peripheral reaction in the sense that it probes mostly the asymptotics of the projectile wave function, i.e.\ its asymptotic normalization constant (ANC) \cite{CN07}.
Therefore, changes in the interior of the wave function, even significant ones, do not seem to affect breakup observables \cite{CN07}.
It is therefore not clear what a fully microscopic description of the projectile, very expensive in a computational viewpoint, would bring to the analysis of breakup reactions.
However, it has also been shown that breakup reactions are sensitive to the description of the continuum of projectile mostly through the phaseshifts \cite{CN06}.
A correct analysis of breakup reactions hence requires a realistic description of the projectile continuum.
As mentioned in \Sec{theory}, the parameters of the core-fragment potential $V_{cf}$ are adjusted to reproduce the binding energy of the system and some of its low-lying states.
However this is usually not sufficient to constrain $V_{cf}$ in all partial waves and in particular to fix their phaseshifts.
Moreover, little---if any---information about this continuum is known experimentally: the core itself is usually radioactive and measuring neutron scattering off it is very difficult---if not impossible.
Being built on first principles, microscopic descriptions of the projectile have more predictive power than phenomenological $c$-$f$ potentials.
They could provide the missing inputs, like phaseshifts, to better constrain $V_{cf}$.

Various techniques could be used to deduce two-body potentials from microscopic models.
One obvious way is to explore the parameter space to reproduce the phaseshifts provided by the microscopic description of the nucleus.
Another way is to build $V_{cf}$ in each partial wave directly from these phaseshifts.
This can be easily done using supersymmetric techniques \cite{BS04}.
The recent developments in halo effective-field theories (EFT) \cite{BHV02,HP11}, open another way to constrain the $c$-$f$ interaction.
The various parameters of this EFT can indeed be fitted to the microscopic-model predictions and hence produce an effective interaction between the core and the fragment that can be used in two-body breakup models.

This would provide a way to include the microscopic inputs that matter in usual reaction models while keeping affordable computational times.
Unfortunately not every kind of reaction can be described in this manner.
A simple two-body description of the projectile is not sufficient to model the case in which there is a significant probably to find the core in one of its excited states.
The breakup of the projectile in three clusters, also known as four-body breakup, can obviously not be reliably modelled using a two-body description for the projectile.
These two cases need special attention.

\subsection{Core excitation}\label{corex}

\begin{figure}[h]
\center
\includegraphics[width=7.5cm]{MLfig2.eps}
\includegraphics[width=7.5cm]{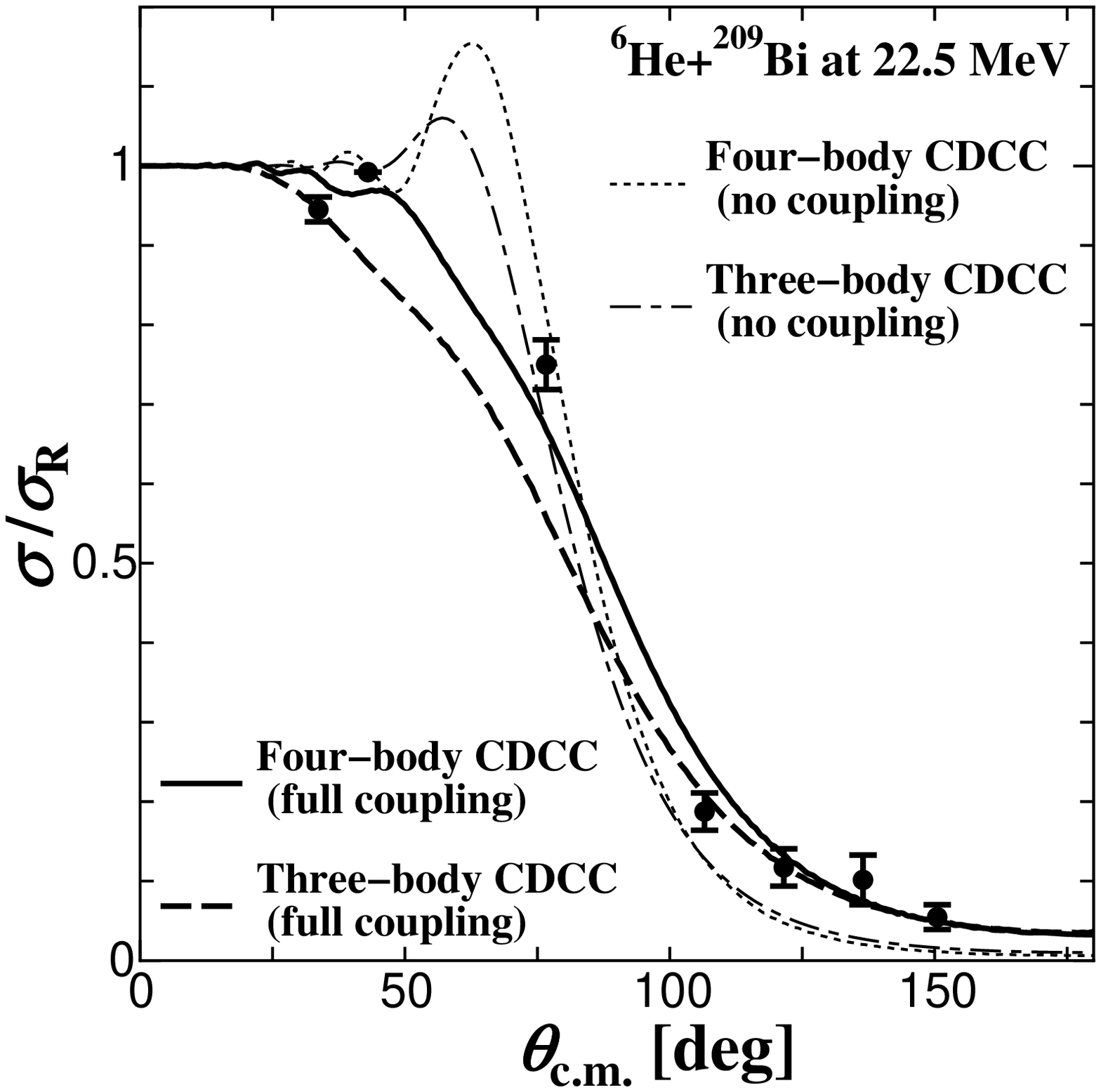}
\caption{Left: Study of the resonant breakup of $^{11}$Be on C at $67A$MeV within a DWBA model including core excitation \cite{ML12} (data from \Ref{Fuk04}). The calculation within a single-particle model (dashed line) and the contribution of the core excitation (dash-dotted line) are shown separately.
Reprinted figure with permission from \Ref{ML12}. Copyright (2012) by the American Physical Society.
Right: CDCC calculations of the elastic scattering of the two-neutron halo nucleus $^6$He off Bi at 22.5~MeV \cite{Mat06} (data from Refs.~\cite{Agu00,Agu01}).
The roles of the three-body structure of the projectile and of the coupling to the breakup channel are illustrated.
Reprinted figure with permission from \Ref{Mat06}. Copyright (2006) by the American Physical Society.}\label{f3}
\end{figure}

The inclusion of the core excitation within the projectile description has first been implemented within the CDCC framework by Summers, Nunes and Thompson \cite{SNT06r,SNT06}.
The calculations performed within this eXtended CDCC model (XCDCC) for $^{11}$Be show little effect of the core excitation on breakup cross sections compared to single-particle calculations \cite{SNT06r,SNT06,SN07}.
However, this seems contradicted by recent DWBA calculations in which a projectile description including core excitation has been implemented \cite{CDM11,MC12,ML12}.
These calculations show significant effects of the core excitation in angular distributions for resonant breakup, which cannot be reproduced with a single-particle model.
This is illustrated in the left panel of \Fig{f3}: the diffractive pattern of the experimental breakup cross sections of $^{11}$Be on C at $67A$MeV \cite{Fuk04} cannot be reproduced considering a single-particle model of the projectile (dashed line) \cite{ML12}.
The contribution of the core excitation (dash-dotted line) and especially its interference with the single-particle breakup plays a crucial role.
This suggests that interesting information about the projectile structure, and in particular its resonant continuum, can be inferred from experimental data using such a model.

Nevertheless, higher-order effects, such as couplings within the projectile continuum, which are neglected in the DWBA framework, may reduce these effects as suggested by XCDCC calculations \cite{SNT06}.
This seems to be confirmed within the new formulation of XCDCC developed by de~Diego \etal \cite{DAL13}.
These results also suggest that the influence of core excitation upon breakup cross sections depends on the considered observable: resonant breakup seems to be more affected than non-resonant breakup.
Further studies of this reaction mechanism will be needed to fully grasp the role played by core excitation in reactions involving halo nuclei.

\subsection{Four-body breakup}\label{4body}
Another critical issue in the description of breakup reactions is the development of a model for the four-body breakup, i.e.\ the breakup of the projectile into three clusters, as it happens for two-nucleon halo nuclei \cite{Aum99,Agu00,Agu01,Nak06,FCR13}.
Obviously this reaction cannot be correctly described assuming a two-body description of the projectile.
Various extensions of the CDCC framework have been developed for three-body projectiles \cite{Mat04,Mat06,Rod08,Rod09}.
The major difficulty lies in handling the large model space required to describe the projectile and in particular its three-body continuum.
In early works the pseudo-state technique has been applied to describe this continuum.
These works have used either a Gaussian basis \cite{Mat04,Mat06} or a THO basis \cite{Rod08} to discretise the continuum.
This leads to tractable calculations but in which the breakup part of the wave function is difficult to analyse: in a pseudo-state it is not clear how the energy is shared between the three fragments.
These calculations were nevertheless able to describe the elastic-scattering of three-body projectiles including a breakup channel.
This shows that both the three-body description of the projectile and the inclusion of the breakup channel are necessary to reproduce elastic-scattering data of $^6$He on various targets and at different beam energies \cite{Mat04,Mat06,Rod08} (see the right panel of \Fig{f3}).
This emphasises the need for including the coupling of different reaction channels in accurate descriptions of collisions involving exotic nuclei (see \Sec{coupling}).

To obtain breakup observables with a better accuracy the binning method has recently been introduced in four-body CDCC calculations \cite{Rod09}.
This method is heavier in a computation point of view than the pseudo-state technique but it provides a finer description of the projectile continuum, which gives more detailed information about the breakup channel.
In particular, total-energy distribution, i.e. the breakup cross section as a function of the total three-body energy in the continuum, is now accessible.

Efforts are also performed to better exploit the calculations involving pseudo-states.
In \Ref{Mat10}, the complex scaling method is coupled to CDCC to smooth the energy distribution obtained from breakup calculations using pseudo-states to describe the continuum.
This technique is shown to work quite well on a three-body test case and leads to a faster convergence and smoother energy distribution than with the binning technique.
Good agreement with experimental data has been obtained for both Coulomb and nuclear-dominated breakups of $^6$He \cite{Mat10}.
More recently, this technique has been extended to obtain double-differential cross sections, i.e.\ Dalitz plots \cite{KMM13}.
These observables show how the energy is distributed among the projectile fragments after the breakup, which provides significant information about the correlation between the projectile constituents \cite{KMM13,KKM10}.
They also emphasise the effects of the interactions between the clusters in the projectile continuum (final-state interaction). As in the two-body case, these interactions are shown to be significant in the calculation of the breakup of Borromean systems \cite{KKM10}.

These results illustrate the significant research activity in this field and also the difficulty to manage a fine description of the continuum for three-body projectiles within the CDCC framework.
Nevertheless, the progresses made recently indicate that we are at the dawn of an era in which a more detailed confrontation between the thee-body structure model of the projectile and experimental data will be possible using four-body CDCC.

Another way to study theoretically the breakup of three-body projectiles is to use a simpler, less computationally intensive,  reaction model than CDCC (see \Sec{reaction}).
In \Ref{BCD09}, the CCE is used to study the Coulomb breakup of $^6$He on Pb at $240A$MeV measured experimentally in \Ref{Aum99}.
Being based on the eikonal model, the CCE requires less computational time to evaluate the breakup cross sections than CDCC.
Thanks to this advantage, a finer description of the core-n-n continuum can be used, which enables an easier calculation of double-differential cross sections.
These results confirm the significance of the final-state interactions in breakup calculations and suggest a slightly dominant $\alpha$-dineutron structure in the $^6$He continuum \cite{BCD09}.
This technique has also been applied for the breakup of $^{11}$Li on Pb at $70A$MeV \cite{Nak06} and fair agreement between theory and experiment has been observed for angular and total-energy distributions \cite{PDB12}.
The calculation of the corresponding double-differential cross section suggests less correlation between both halo neutrons in $^{11}$Li continuum than in $^6$He.
Unfortunately, no experimental double-differential cross section is available yet, which would enable to constrain theoretical models of Borromean nuclei.
Significant efforts, both theoretical and experimental, should be made in order to obtain such observables as they seem to convey interesting information about the structure of two-neutron halo nuclei.

Albeit interesting for its low computational cost, the CCE remains a simple reaction model, valid only at sufficiently high energy, and which neglects part of the reaction dynamics \cite{CBS08}.
In this, CDCC is superior since it is valid at all beam energies and includes all couplings at all orders.
As one can see here a compromise has still to be made between the quality of the projectile description and the accuracy of the reaction model.
Maybe an intermediate solution could be found by developing a four-body breakup model within the E-CDCC or DEA frameworks.
These models are more computationally effective than CDCC, while including sufficient projectile dynamics to reproduce CDCC calculations at sufficiently high energy \cite{CEN12,FOC14}.
They are thus interesting alternatives to the full CDCC and the simple CCE to describe reactions involving three-body nuclei.
Hopefully both requirements of an accurate reaction model coupled to a precise projectile description will be met within one model of four-body breakup in a near future.

\section{Reaction modelling}\label{reaction}
Another key point to improve the quality of the structure information inferred from breakup experiments is the description of the reaction process itself.
The first analyses of breakup reactions were performed using perturbative models, in which the transition from the initial bound state to the projectile continuum takes place in one single step.
The development of more accurate reaction models has shown that the reaction mechanism is more complex than this simple transition, and that couplings inside the continuum affect breakup observables.
As explained in Refs.~\cite{EB96,EBS05,CB05,GCB07}, this may lead to misinterpretation of experimental data (see also Secs.~\ref{corex} and \ref{4body} and \Ref{Tho14}).

Besides being able to describe accurately the reaction mechanisms, breakup models should also be numerically tractable.
As mentioned previously, it is challenging to include within CDCC descriptions of the projectile that go beyond the simple two-body model.
This is due to the huge model space required to reliably include finer descriptions of the projectile, aggravated by the heaviness of the CDCC calculations.
Using simpler approximations may help keeping computational times affordable without necessarily loosing predictive power in the reaction modelling \cite{CEN12}.
However, the validity range of these approximations must be well established, e.g.\ using other models and/or precise experimental data for validation.

\subsection{Coupling breakup models to other reaction processes}\label{coupling}
Recent experiments have confirmed the complexity of the reaction process.
Measuring the elastic scattering of $^{9,10,11}$Be isotopes on Zn around the Coulomb barrier \cite{Dip10}, Di~Pietro \etal have seen that the results obtained for $^{11}$Be
differ significantly from those for the other beryllium isotopes: the $^{11}$Be elastic-scattering cross section cannot be reproduced using a simple folding procedure, even if its halo structure is taken into account.
The confrontation with CDCC calculations has shown that this is due to a significant coupling between the elastic-scattering channel and breakup \cite{Dip10,Dip12}, as already noted in four-body CDCC calculations \cite{Mat04,Mat06,Rod08} and in a theoretical analysis of angular distributions \cite{CHB10}.
This strong coupling between these two processes indicates that the description of a reaction involving exotic nuclei may have to encompass more than the sole mechanism in which one is interested.

Another example in which the analysis of experimental data is at stake is the one-nucleon knockout reaction (KO) \cite{Sau00}.
In that reaction one nucleon is removed from the projectile through its interaction with a light target (e.g.\ C or Be).
The KO cross sections are inclusive in the sense that the removed nucleon is not measured in coincidence with the core.
This reaction thus includes not only the deeply inelastic removal of the nucleon, but also the elastic breakup channel described here.
A coherent description of both processes is therefore needed for a reliable analysis of the data.

KO experiments are usually performed to extract spectroscopic factors \cite{HT03}.
This is done confronting experimental cross sections with eikonal calculations of the reaction that use shell-model results as structure inputs.
Surprisingly, this confrontation leads to a systematic reduction of the measured cross sections relative to their theoretical predictions \cite{Gad08}.
Part of the problem comes from the description of the nuclear structure because the shell model cannot account for the coupling with the nucleus spectrum above the nucleon-emission threshold, which is significant for loosely-bound nuclei \cite{JHH11}.
However, it seems also that part of the problem is due to the use of the sudden approximation in the reaction modelling \cite{FOB12}.
To test this hypothesis, the role played by the sudden approximation in the description of KO should be evaluated.
Since E-CDCC and DEA are eikonal-based models, which do not include the sudden approximation, they could be extended to study KO reactions.
This has been suggested by Yahiro \etal for the E-CDCC model \cite{YOM11}.
However no differential cross section has been obtained within this new model so that the hypothesis of \Ref{FOB12} has not yet been tested.
The development of new KO models that include the projectile dynamics and that can describe elastic breakup and KO on the same footing may hence help to solve the long-standing problem summarised in \Ref{Gad08}.

\subsection{Extending the range of validity of existing models of breakup}\label{range}
As mentioned in \Sec{structure}, it is computationally challenging to include within CDCC descriptions of the projectile that go beyond the simple two-body model.
Simpler reaction models may enable the use of more realistic descriptions of the projectile, while keeping computational times affordable.
For example, in Refs.~\cite{CDM11,MC12,ML12}, DWBA-based models have been used to analyse the influence of core excitation upon breakup observables (see \Sec{corex}).
In Refs.~\cite{BCD09,PDB12}, the CCE is used to describe the breakup of the two-neutron halo nuclei $^6$He and $^{11}$Li (see \Sec{4body}).
In both cases, these analyses have revealed interesting effects of the projectile structure upon breakup observables, which are difficult to access to full CDCC calculations.
These models are thus an interesting way to explore the influence of the projectile structure on reaction data.

The eikonal-based reaction models E-CDCC \cite{OYI03} and DEA \cite{BCG05} are particularly interesting for such an extension.
First, they are less time-consuming than full CDCC calculations, while providing reliable theoretical cross sections \cite{CEN12,FOC14}.
Second, they include the dynamical effects missing in DWBA and CCE, which may play a significant effect on breakup observables (see Secs.~\ref{corex} and \ref{4body}).
Unfortunately, E-CDCC and DEA are limited to intermediate/high energies and therefore cannot be applied for ISOL-type experiments.
As shown in \Ref{CEN12}, the major problem of these approximations is the lack of Coulomb deflection.
They indeed assume that the projectile-target relative motion is not much different from the incoming plane wave (see \Sec{theory}).
This assumption is no longer valid at low energy, at which the projectile is significantly deflected by the Coulomb field of the target.
An inclusion of the Coulomb deflection would help increasing the range of validity of eikonal-like models down to lower energies.
This would help analysing low-energy experiments without the need of heavy CDCC calculations.

For example, being based on an expansion of the wave function identical to that of CDCC (see \Sec{theory}), E-CDCC can be improved readily into a hybrid version \cite{OYI03} in which the $P$-$T$ relative angular momenta $L$ are separated into two distinct regions.
Low $L$s are treated exactly within CDCC, i.e.\ including all possible couplings, whereas the simpler eikonal approximation is used to compute larger $L$s.
This enables to include the Coulomb deflection and hence obtain results identical to those of CDCC within a shorter computational time \cite{FOC14}.

On the other side of the energy range, a problem that has not yet been fully investigated is that of the relativistic effects.
If one excepts an extension of the first-order perturbation theory of Alder and Winther \cite{AW75}, breakup models have mostly been developed within non-relativistic quantum mechanics.
In a first analysis of relativistic effects in breakup reactions, Bertulani indicates that these effects play a role, even at intermediate energies \cite{Ber05}.
In Refs.~\cite{OB09,OB10} Ogata and Bertulani study these effects in more details using a relativistic correction to E-CDCC.
They show that this correction affects mostly the Coulomb $P$-$T$ interaction, which leads to changes in the breakup cross section at forward angles.
These changes are of the order of 10\% at $100A$MeV and 15\% at $250A$MeV.
Due to the strong non-linearity of the Coulomb couplings to and within the continuum, no simple correction could be found to simulate relativistic effects in cross sections obtained from non-relativistic reaction models.
Therefore, relativistic models seem to be unavoidable to accurately analyse data obtained at large beam energies, i.e.\ above 100$A$MeV.
With the increase of beam energies at various RIB facilities, these effects should be taken into account for a proper study of nuclear structure from breakup measurements.

\section{New reaction observables}\label{observable}

An alternative to developing complex models to describe accurately the reaction mechanism is to search for observables that are not sensitive to that mechanism.
In this way only a simple reaction model is necessary to analyse experimental data.
Moreover, such observables would emphasise information about the projectile structure as it would not be hidden by reaction artefacts.

\begin{figure}
\center
\includegraphics[width=8cm]{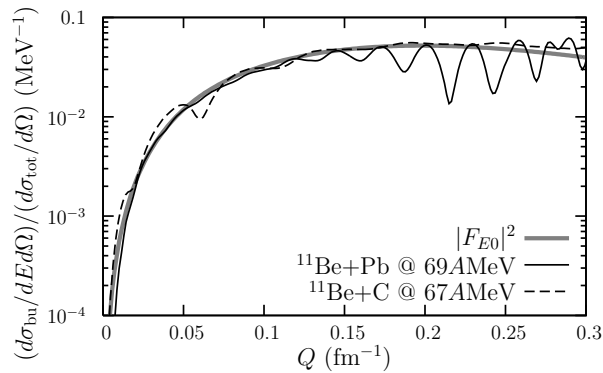}
\caption{The ratio evaluated for $^{11}$Be impinging on Pb at $69A$MeV (solid line) and C at $67A$MeV (dashed line) are shown to be very similar.
They are also in excellent agreement with an adiabatic reaction model (thick grey line) \cite{CJN11}.
Reprinted figure from \Ref{CJN11}, Copyright (2011), with permission from Elsevier.}\label{f4}
\end{figure}

The \emph{ratio method} is such a new reaction observable \cite{CJN11,CJN13}.
It consists of the ratio between two angular distributions for two different processes, e.g.\ breakup and elastic scattering.
As shown in \Ref{CHB10}, both angular distributions exhibit similar patterns reflecting the way the projectile is scattered off the target: Coulomb rainbow, Near/Far interferences etc.
Taking their ratio removes most of this angular dependence, leading to an observable nearly independent of the reaction process.
Calculations within the DEA show that, for one-neutron halo nuclei, the ratio is nearly the same for light and heavy targets, confirming its independence to the reaction mechanism (see Refs.~\cite{CJN11,CJN13} and \Fig{f4}).

Thanks to this independence, the ratio is strongly sensitive to the structure of the projectile.
In particular it provides precise information about the binding energy of the halo neutron and its partial wave.
Depending on the scattering angle, the ratio probes different parts of the radial wave function of the projectile:
the ANC at forward angles, and, unlike most reaction observables, the internal part at larger angles \cite{CJN13}.

Up to now this ratio has been developed only for a simple two-body description of one-neutron halo projectiles.
Preliminary calculations suggest that charged cases, such as proton haloes, and three-body projectiles, such as two-neutron halo nuclei, can also be studied with this method.
Although there are not yet data on which to apply the ratio method, the breakup probability measured in \Ref{FCR13} for the collision of $^{11}$Li on Pb around the Coulomb barrier shows great similarities with one of the ratios suggested in \Ref{CJN13}.
This seems to confirm the possibility to extend the ratio technique to Borromean nuclei and to low energies.
It would also be interesting to see whether this idea can be extended to other reactions, such as transfer.
This would certainly help analysing data in a more model-independent way.

\section{Outlook}\label{conclusion}
Breakup reaction is one of the best tools to study the structure of halo nuclei.
Various models have been developed to describe the reaction process and infer information about this exotic nuclear structure: CDCC \cite{Kam86,Yah86,TNT01}, TD \cite{KYS94,EBB95,TW99,CBM03c,LSC99}, E-CDCC \cite{OYI03}, DEA \cite{BCG05}, CCE \cite{MBB03} (see also \Ref{BC12} for a recent review).
Thanks to intensive studies of breakup, the reaction mechanism is now rather well understood.
However, these studies have shown the need of an accurate description of the reaction mechanism as significant higher-order effects take place during the collision, which may affect the quality of the structure information inferred from experimental data \cite{EB96,EBS05,CB05,GCB07}.

Most of the current reaction models are limited to a simple two-body description of the projectile.
Due to the peripheral nature of breakup reactions, it is probably not necessary to include a fully microscopic description of the projectile within accurate reaction models. Nevertheless, microscopic structure calculations could provide important information to constrain phenomenological core-fragment potentials.
To go beyond this simple model, various efforts have been made to extend the CDCC framework to descriptions of the projectile in which the core can be in an excited state \cite{SNT06r,SNT06,DAL13} or for three-body projectiles \cite{Mat04,Mat06,Rod08,Rod09,Mat10}.
Unfortunately these extensions are computationally challenging and provide mostly inclusive breakup observables.
To obtain more differential observables one has to rely on simpler models, such as DWBA \cite{CDM11,MC12,ML12} or CCE \cite{BCD09,PDB12}.
These studies suggest that interesting information can be inferred from differential cross sections.
They also indicate that a full CDCC model might not be necessary to analyse experimental data and that simpler descriptions of breakup, such as E-CDCC or DEA, may be used, as long as their domain of validity is well under control \cite{CEN12}.
Possible extensions of these domains of validity, e.g.\ to lower beam energies, would help analysing data without having to resort to full CDCC calculations.

An alternative to the development of accurate reaction models is the search for reaction observables that are independent of the reaction process.
One example of such observable is the ratio of angular distributions \cite{CJN11}, which, thanks to its independence to the reaction mechanism, does not require precise models for its analysis.
It is also more sensitive than other observables to the projectile structure \cite{CJN13}.

The aforementioned studies show how complicated the description of reactions involving exotic nuclei can be, not only because of the intrinsic complexity of the reaction mechanism, but also because of the coupling it can have with other processes.
The development of new RIB facilities such as RIBF in Japan, FRIB in the USA or FAIR in Europe will improve the rate at which exotic nuclei can be produced and hence the statistics in reaction measurements.
This will require precise theoretical reaction models that can account for the aforementioned effects in order to extract valuable structure information from future data.

\section*{Acknowledgments}
This text presents research results
of the Belgian Research Initiative on eXotic nuclei (BriX),
Program No. P7/12, on interuniversity attraction poles of the
Belgian Federal Science Policy Office.

\section*{References}

\begin{thebibliography}{81}

\bibitem{Tan85r}
I.~Tanihata, H.~Hamagaki, O.~Hashimoto, Y.~Shida, N~Yoshikawa, K.~Sugimoto,
  O.~Yamakawa, T.~Kobayashi, and N.~Takahashi.
\newblock Measurements of interaction cross sections and nuclear radii in the
  light $p$-shell region.
\newblock {\em Phys. Rev. Lett.}, 55:2676, 1985.

\bibitem{Tan85b}
I.~Tanihata, H.~Hamagaki, O.~Hashimoto, S.~Nagamiya, Y.~Shida, N~Yoshikawa,
  O.~Yamakawa, K.~Sugimoto, T.~Kobayashi, D.~E. Greiner, N.~Takahashi, and
  Y.~Nojiri.
\newblock Measurements of interaction cross sections and radii of {He}
  isotopes.
\newblock {\em Phys. Lett. B}, 160:380, 1985.

\bibitem{WBB90}
E.~K. Warburton, J.~A. Becker, and B.~A. Brown.
\newblock Mass systematics for ${A}=29$--44 nuclei: The deformed ${A}\sim32$
  region.
\newblock {\em Phys. Rev. C}, 41:1147--1166, 1990.

\bibitem{Tan96}
I.~Tanihata.
\newblock Neutron halo nuclei.
\newblock {\em J. Phys. G}, 22:157, 1996.

\bibitem{Jon04}
B.~Jonson.
\newblock Light dripline nuclei.
\newblock {\em Phys. Rep.}, 389:1, 2004.

\bibitem{HJ87}
P.~G. Hansen and B.~Jonson.
\newblock The neutron halo of extremely neutron-rich nuclei.
\newblock {\em Europhys. Lett.}, 4:409, 1987.

\bibitem{Dip10}
A.~Di~Pietro, G.~Randisi, V.~Scuderi, L.~Acosta, F.~Amorini, M.~J.~G. Borge,
  P.~Figuera, M.~Fisichella, L.~M. Fraile, J.~Gomez-Camacho, H.~Jeppesen,
  M.~Lattuada, I.~Martel, M.~Milin, A.~Musumarra, M.~Papa, M.~G. Pellegriti,
  F.~Perez-Bernal, R.~Raabe, F.~Rizzo, D.~Santonocito, G.~Scalia, O.~Tengblad,
  D.~Torresi, A.~Maira Vidal, D.~Voulot, F.~Wenander, and M.~Zadro.
\newblock Elastic scattering and reaction mechanisms of the halo nucleus
  $^{11}${Be} around the {C}oulomb barrier.
\newblock {\em Phys. Rev. Lett.}, 105:022701, 2010.

\bibitem{Fuk04}
N.~Fukuda, T.~Nakamura, N.~Aoi, N.~Imai, M.~Ishihara, T.~Kobayashi, H.~Iwasaki,
  T.~Kubo, A.~Mengoni, M~Notani, H.~Otsu, H.~Sakurai, S.~Shimoura,
  T.~Teranishi, Y.~X. Watanabe, and K.~Yoneda.
\newblock Coulomb and nuclear breakup of a halo nucleus {$^{11}$\rm Be}.
\newblock {\em Phys. Rev. C}, 70:054606, 2004.

\bibitem{Sau00}
E.~Sauvan, F.~Carstoiu, N.A. Orr, J.C. Ang\'elique, W.N. Catford, N.M. Clarke,
  M.~Mac Cormick, N.~Curtis, M.~Freer, S.~Gr\'evy, C.~Le Brun, M.~Lewitowicz,
  E.~Li\'egard, F.M. Marqu\'es, P.~Roussel-Chomaz, M.G.~Saint Laurent,
  M.~Shawcross, and J.S. Winfield.
\newblock One-neutron removal reactions on neutron-rich psd-shell nuclei.
\newblock {\em Phys. Lett. B}, 491:1, 2000.

\bibitem{Jon10}
K.~L. Jones, A.~S. Adekola, D.~W. Bardayan, J.~C. Blackmon, K.~Y. Chae, K.~A.
  Chipps, J.~A. Cizewski, L.~Erikson, C.~Harlin, R.~Hatarik, R.~Kapler, R.~L.
  Kozub, J.~F. Liang, R.~Livesay, Z.~Ma, B.~H. Moazen, C.~D. Nesaraja, F.~M.
  Nunes, S.~D. Pain, N.~P. Patterson, D.~Shapira, J.~F. Shriner, M.~S. Smith,
  T.~P. Swan, and J.~S. Thomas.
\newblock The magic nature of $^{132}${Sn} explored through the single-particle
  states of $^{133}${Sn}.
\newblock {\em Nature}, 465:454, 2010.

\bibitem{Kam86}
M.~Kamimura, M.~Yahiro, Y.~Iseri, Y.~Sakuragi, H.~Kameyama, and M.~Kawai.
\newblock Projectile breakup processes in nuclear reactions.
\newblock {\em Prog. Theor. Phys. Suppl.}, 89:1, 1986.

\bibitem{Yah86}
M.~Yahiro, Y.~Iseri, H.~Kameyama, M.~Kamimura, and M.~Kawai.
\newblock Effects of deuteron virtual breakup on deuteron elastic and inelastic
  scattering.
\newblock {\em Prog. Theor. Phys. Suppl.}, 89:32, 1986.

\bibitem{TNT01}
J.~A. Tostevin, F.~M. Nunes, and I.~J. Thompson.
\newblock Calculations of three-body observables in {$^8$\rm B} breakup.
\newblock {\em Phys. Rev. C}, 63:024617, 2001.

\bibitem{KYS94}
T.~Kido, K.~Yabana, and Y.~Suzuki.
\newblock Coulomb breakup mechanism of neutron drip-line nuclei.
\newblock {\em Phys. Rev. C}, 50:R1276, 1994.

\bibitem{EBB95}
H.~Esbensen, G.~F. Bertsch, and C.~A. Bertulani.
\newblock Higher-order dynamical effects in {C}oulomb dissociation.
\newblock {\em Nucl. Phys. A}, 581:107, 1995.

\bibitem{TW99}
S.~Typel and H.~H. Wolter.
\newblock Dynamical description of {C}oulomb dissociation.
\newblock {\em Z. Naturforsch. Teil A}, 54:63, 1999.

\bibitem{CBM03c}
P.~Capel, D.~Baye, and V.~S. Melezhik.
\newblock Time-dependent analysis of the breakup of halo nuclei.
\newblock {\em Phys. Rev. C}, 68:014612, 2003.

\bibitem{LSC99}
D.~Lacroix, J.~A. Scarpaci, and Ph. Chomaz.
\newblock Theoretical description of the towing mode through a time-dependent
  quantum calculation.
\newblock {\em Nuclear Physics A}, 658:273, 1999.

\bibitem{Glauber}
R.~J. Glauber.
\newblock High energy collision theory.
\newblock In W.~E. Brittin and L.~G. Dunham, editors, {\em Lecture in
  Theoretical Physics}, volume~1, page 315. Interscience, New York, 1959.

\bibitem{OYI03}
K.~Ogata, M.~Yahiro, Y.~Iseri, T.~Matsumoto, and M.~Kamimura.
\newblock New coupled-channel approach to nuclear and {C}oulomb breakup
  reactions.
\newblock {\em Phys. Rev. C}, 68:064609, 2003.

\bibitem{BCG05}
D.~Baye, P.~Capel, and G.~Goldstein.
\newblock Collisions of halo nuclei within a dynamical eikonal approximation.
\newblock {\em Phys. Rev. Lett.}, 95:082502, 2005.

\bibitem{MBB03}
J.~Margueron, A.~Bonaccorso, and D.~M. Brink.
\newblock A non-perturbative approach to halo breakup.
\newblock {\em Nucl. Phys. A}, 720:337, 2003.

\bibitem{AS04}
B.~Abu-Ibrahim and Y.~Suzuki.
\newblock Breakup of one-neutron halo nuclei within eikonal model.
\newblock {\em Prog. Theor. Phys.}, 112:1013, 2004.

\bibitem{CBS08}
P.~Capel, D.~Baye, and Y.~Suzuki.
\newblock Coulomb-corrected eikonal description of the breakup of halo nuclei.
\newblock {\em Phys. Rev. C}, 78:054602, 2008.

\bibitem{BC12}
D.~Baye and P.~Capel.
\newblock Breakup reaction models for two- and three-cluster projectiles.
\newblock In Christian Beck, editor, {\em Clusters in Nuclei, Vol.2}, volume
  848 of {\em Lecture Notes in Physics}, pages 121--163. Springer Berlin
  Heidelberg, 2012.

\bibitem{CJN11}
P.~Capel, R.C. Johnson, and F.M. Nunes.
\newblock One-neutron halo structure by the ratio method.
\newblock {\em Phys. Lett. B}, 705:112, 2011.

\bibitem{DFS05}
A.~Deltuva, A.~C. Fonseca, and P.~U. Sauer.
\newblock Momentum-space treatment of the {C}oulomb interaction in
  three-nucleon reactions with two protons.
\newblock {\em Phys. Rev. C}, 71:054005, 2005.

\bibitem{DFS05b}
A.~Deltuva, A.~C. Fonseca, and P.~U. Sauer.
\newblock Momentum-space description of three-nucleon breakup reactions
  including the {C}oulomb interaction.
\newblock {\em Phys. Rev. C}, 72:054004, 2005.

\bibitem{Kaw86}
M.~Kawai.
\newblock Formalism of the method of coupled discretized continuum channels.
\newblock {\em Prog. Theor. Phys. Suppl.}, 89:11, 1986.

\bibitem{AYK89}
N.~Austern, M.~Yahiro, and M.~Kawai.
\newblock Continuum discretized coupled-channels method as a truncation of a
  connected-kernel formulation of three-body problems.
\newblock {\em Phys. Rev. Lett.}, 63:2649, 1989.

\bibitem{AKY96}
N.~Austern, M.~Kawai, and M.~Yahiro.
\newblock Three-body reaction theory in a model space.
\newblock {\em Phys. Rev. C}, 53:314, 1996.

\bibitem{CEN12}
P.~Capel, H.~Esbensen, and F.~M. Nunes.
\newblock Comparing nonperturbative models of the breakup of neutron-halo
  nuclei.
\newblock {\em Phys. Rev. C}, 85:044604, 2012.

\bibitem{Tho14}
I.~Thompson.
\newblock Computational challenges in theories of nuclear reactions.
\newblock {\em J. Phys. G}, **:***, 2014.

\bibitem{Mat03}
T.~Matsumoto, T.~Kamizato, K.~Ogata, Y.~Iseri, E.~Hiyama, M.~Kamimura, and
  M.~Yahiro.
\newblock New treatment of breakup continuum in the method of continuum
  discretized coupled channels.
\newblock {\em Phys. Rev. C}, 68:064607, 2003.

\bibitem{PBM01}
F.~P\'erez-Bernal, I.~Martel, J.~M. Arias, and J.~G\'omez-Camacho.
\newblock Continuum discretization in a basis of transformed
  harmonic-oscillator states.
\newblock {\em Phys. Rev. A}, 63:052111, 2001.

\bibitem{DBD10}
T.~Druet, D.~Baye, P.~Descouvemont, and J.-M. Sparenberg.
\newblock {CDCC} calculations with the {L}agrange-mesh technique.
\newblock {\em Nucl. Phys. A}, 845:88, 2010.

\bibitem{AW75}
K.~Alder and A.~Winther.
\newblock {\em Electromagnetic Excitation}.
\newblock North-Holland, Amsterdam, 1975.

\bibitem{FOC14}
T.~Fukui, K.~Ogata, and P.~Capel.
\newblock Analysis of a low-energy correction to the eikonal approximation,
  2014.
\newblock Submitted for publication.

\bibitem{GBC06}
G.~Goldstein, D.~Baye, and P.~Capel.
\newblock Dynamical eikonal approximation in breakup reactions of {$^{11}$\rm
  Be}.
\newblock {\em Phys. Rev. C}, 73:024602, 2006.

\bibitem{BCD09}
D.~Baye, P.~Capel, P.~Descouvemont, and Y.~Suzuki.
\newblock Four-body calculation of {$^6$\rm He} breakup with the
  {C}oulomb-corrected eikonal method.
\newblock {\em Phys. Rev. C}, 79:024607, 2009.

\bibitem{PDB12}
E.~C. Pinilla, P.~Descouvemont, and D.~Baye.
\newblock Three-body breakup of $^{11}${Li} with the eikonal method.
\newblock {\em Phys. Rev. C}, 85:054610, 2012.

\bibitem{DH13}
P.~Descouvemont and M.~S. Hussein.
\newblock Towards a microscopic description of reactions involving exotic
  nuclei.
\newblock {\em Phys. Rev. Lett.}, 111:082701, 2013.

\bibitem{CN07}
P.~Capel and F.~M. Nunes.
\newblock Peripherality of breakup reactions.
\newblock {\em Phys. Rev. C}, 75:054609, 2007.

\bibitem{CN06}
P.~Capel and F.~M. Nunes.
\newblock Influence of the projectile description on breakup calculations.
\newblock {\em Phys. Rev. C}, 73:014615, 2006.

\bibitem{BS04}
D.~Baye and J.-M. Sparenberg.
\newblock Inverse scattering with supersymmetric quantum mechanics.
\newblock {\em J. Phys. A}, 37:10223, 2004.

\bibitem{BHV02}
C.A. Bertulani, H.-W. Hammer, and U.~van Kolck.
\newblock Effective field theory for halo nuclei: shallow p-wave states.
\newblock {\em Nucl. Phys. A}, 712:37, 2002.

\bibitem{HP11}
H.-W. Hammer and D.R. Phillips.
\newblock Electric properties of the {B}eryllium-11 system in halo {EFT}.
\newblock {\em Nucl. Phys. A}, 865:17, 2011.

\bibitem{SNT06r}
N.~C. Summers, F.~M. Nunes, and I.~J. Thompson.
\newblock Core transitions in the breakup of exotic nuclei.
\newblock {\em Phys. Rev. C}, 73:031603, 2006.

\bibitem{SNT06}
N.~C. Summers, F.~M. Nunes, and I.~J. Thompson.
\newblock Extended continuum discretized coupled channels method: Core
  excitation in the breakup of exotic nuclei.
\newblock {\em Phys. Rev. C}, 74:014606, 2006.

\bibitem{SN07}
N.~C. Summers and F.~M. Nunes.
\newblock Core excitation in the elastic scattering and breakup of $^{11}${Be}
  on protons.
\newblock {\em Phys. Rev. C}, 76:014611, 2007.

\bibitem{CDM11}
R.~Crespo, A.~Deltuva, and A.~M. Moro.
\newblock Core excitation contributions to the breakup of the one-neutron halo
  nucleus $^{11}${Be} on a proton.
\newblock {\em Phys. Rev. C}, 83:044622, 2011.

\bibitem{MC12}
A.~M. Moro and R.~Crespo.
\newblock Core excitation effects in the breakup of the one-neutron halo
  nucleus $^{11}${Be} on a proton target.
\newblock {\em Phys. Rev. C}, 85:054613, 2012.

\bibitem{ML12}
A.~M. Moro and J.~A. Lay.
\newblock Interplay between valence and core excitation mechanisms in the
  breakup of halo nuclei.
\newblock {\em Phys. Rev. Lett.}, 109:232502, 2012.

\bibitem{DAL13}
R.~de~Diego, J.~M. Arias, J.~A. Lay, and A.~M. Moro.
\newblock Continuum-discretized coupled-channels calculations with core
  excitation, 2013.
\newblock arXiv:1312.5684.

\bibitem{Aum99}
T.~Aumann, D.~Aleksandrov, L.~Axelsson, T.~Baumann, M.~J.~G. Borge, L.~V.
  Chulkov, J.~Cub, W.~Dostal, B.~Eberlein, Th.~W. Elze, H.~Emling, H.~Geissel,
  V.~Z. Goldberg, M.~Golovkov, A.~Gr\"unschlo\ss{}, M.~Hellstr\"om, K.~Hencken,
  J.~Holeczek, R.~Holzmann, B.~Jonson, A.~A. Korshenninikov, J.~V. Kratz,
  G.~Kraus, R.~Kulessa, Y.~Leifels, A.~Leistenschneider, T.~Leth, I.~Mukha,
  G.~M\"unzenberg, F.~Nickel, T.~Nilsson, G.~Nyman, B.~Petersen, M.~Pf\"utzner,
  A.~Richter, K.~Riisager, C.~Scheidenberger, G.~Schrieder, W.~Schwab,
  H.~Simon, M.~H. Smedberg, M.~Steiner, J.~Stroth, A.~Surowiec, T.~Suzuki,
  O.~Tengblad, and M.~V. Zhukov.
\newblock Continuum excitations in $^6${He}.
\newblock {\em Phys. Rev. C}, 59:1252, 1999.

\bibitem{Agu00}
E.~F. Aguilera, J.~J. Kolata, F.~M. Nunes, F.~D. Becchetti, P.~A. DeYoung,
  M.~Goupell, V.~Guimar\~aes, B.~Hughey, M.~Y. Lee, D.~Lizcano,
  E.~Martinez-Quiroz, A.~Nowlin, T.~W. O'Donnell, G.~F. Peaslee, D.~Peterson,
  P.~Santi, and R.~White-Stevens.
\newblock Transfer and/or breakup modes in the $^6${He}+$^{209}${Bi} reaction
  near the {C}oulomb barrier.
\newblock {\em Phys. Rev. Lett.}, 84:5058, 2000.

\bibitem{Agu01}
E.~F. Aguilera, J.~J. Kolata, F.~D. Becchetti, P.~A. DeYoung, J.~D. Hinnefeld,
  \'A. Horv\'ath, L.~O. Lamm, Hye-Young Lee, D.~Lizcano, E.~Martinez-Quiroz,
  P.~Mohr, T.~W. O'Donnell, D.~A. Roberts, and G.~Rogachev.
\newblock Elastic scattering and transfer in the system $^6${He}+$^{209}${Bi}
  below the {C}oulomb barrier.
\newblock {\em Phys. Rev. C}, 63:061603, 2001.

\bibitem{Nak06}
T.~Nakamura, A.~M. Vinodkumar, T.~Sugimoto, N.~Aoi, H.~Baba, D.~Bazin,
  N.~Fukuda, T.~Gomi, H.~Hasegawa, N.~Imai, M.~Ishihara, T.~Kobayashi,
  Y.~Kondo, T.~Kubo, M.~Miura, T.~Motobayashi, H.~Otsu, A.~Saito, H.~Sakurai,
  S.~Shimoura, K.~Watanabe, Y.~X. Watanabe, T.~Yakushiji, Y.~Yanagisawa, and
  K.~Yoneda.
\newblock Observation of strong low-lying {E1} strength in the two-neutron halo
  nucleus $^{11}${Li}.
\newblock {\em Phys. Rev. Lett.}, 96:252502, 2006.

\bibitem{FCR13}
J.~P. Fern\'andez-Garc\'{\i}a, M.~Cubero, M.~Rodr\'{\i}guez-Gallardo,
  L.~Acosta, M.~Alcorta, M.~A.~G. Alvarez, M.~J.~G. Borge, L.~Buchmann, C.~A.
  Diget, H.~A. Falou, B.~R. Fulton, H.~O.~U. Fynbo, D.~Galaviz,
  J.~G\'omez-Camacho, R.~Kanungo, J.~A. Lay, M.~Madurga, I.~Martel, A.~M. Moro,
  I.~Mukha, T.~Nilsson, A.~M. S\'anchez-Ben\'{\i}tez, A.~Shotter, O.~Tengblad,
  and P.~Walden.
\newblock $^{11}${Li} breakup on $^{208}${Pb} at energies around the {C}oulomb
  barrier.
\newblock {\em Phys. Rev. Lett.}, 110:142701, 2013.

\bibitem{Mat04}
T.~Matsumoto, E.~Hiyama, K.~Ogata, Y.~Iseri, M.~Kamimura, S.~Chiba, and
  M.~Yahiro.
\newblock Continuum-discretized coupled-channels method for four-body nuclear
  breakup in $^6${He}+$^{12}${C} scattering.
\newblock {\em Phys. Rev. C}, 70:061601(R), 2004.

\bibitem{Mat06}
T.~Matsumoto, T.~Egami, K.~Ogata, Y.~Iseri, M.~Kamimura, and M.~Yahiro.
\newblock Coulomb breakup effects on the elastic cross section of
  $^6${He}+$^{209}${Bi} scattering near {C}oulomb barrier energies.
\newblock {\em Phys. Rev. C}, 73:051602, 2006.

\bibitem{Rod08}
M.~Rodr\'{\i}guez-Gallardo, J.~M. Arias, J.~G\'omez-Camacho, R.~C. Johnson,
  A.~M. Moro, I.~J. Thompson, and J.~A. Tostevin.
\newblock Four-body continuum-discretized coupled-channels calculations using a
  transformed harmonic oscillator basis.
\newblock {\em Phys. Rev. C}, 77:064609, 2008.

\bibitem{Rod09}
M.~Rodr\'{\i}guez-Gallardo, J.~M. Arias, J.~G\'omez-Camacho, A.~M. Moro, I.~J.
  Thompson, and J.~A. Tostevin.
\newblock Four-body continuum-discretized coupled-channels calculations.
\newblock {\em Phys. Rev. C}, 80:051601, 2009.

\bibitem{Mat10}
T.~Matsumoto, K.~Kat\={o}, and M.~Yahiro.
\newblock New description of the four-body breakup reaction.
\newblock {\em Phys. Rev. C}, 82:051602, 2010.

\bibitem{KMM13}
Y.~Kikuchi, T.~Matsumoto, K.~Minomo, and K.~Ogata.
\newblock Two neutron decay from the $2^+_1$ state of $^6${He}.
\newblock {\em Phys. Rev. C}, 88:021602, 2013.

\bibitem{KKM10}
Y.~Kikuchi, K.~Kat\ifmmode~\bar{o}\else \={o}\fi{}, T.~Myo, M.~Takashina, and
  K.~Ikeda.
\newblock Two-neutron correlations in $^6${He} in a {C}oulomb breakup reaction.
\newblock {\em Phys. Rev. C}, 81:044308, 2010.

\bibitem{EB96}
H.~Esbensen and G.F. Bertsch.
\newblock Effects of {E2} transitions in the {C}oulomb dissociation of $^8${B}.
\newblock {\em Nucl. Phys. A}, 600:37, 1996.

\bibitem{EBS05}
H.~Esbensen, G.~F. Bertsch, and K.~A. Snover.
\newblock Reconciling coulomb dissociation and radiative capture measurements.
\newblock {\em Phys. Rev. Lett.}, 94:042502, 2005.

\bibitem{CB05}
P.~Capel and D.~Baye.
\newblock Coupling-in-the-continuum effects in {C}oulomb dissociation of halo
  nuclei.
\newblock {\em Phys. Rev. C}, 71:044609, 2005.

\bibitem{GCB07}
G.~Goldstein, P.~Capel, and D.~Baye.
\newblock Analysis of {C}oulomb breakup experiments of {$^8$\rm B} with a
  dynamical eikonal approximation.
\newblock {\em Phys. Rev. C}, 76:024608, 2007.

\bibitem{Dip12}
A.~Di~Pietro, V.~Scuderi, A.~M. Moro, L.~Acosta, F.~Amorini, M.~J.~G. Borge,
  P.~Figuera, M.~Fisichella, L.~M. Fraile, J.~Gomez-Camacho, H.~Jeppesen,
  M.~Lattuada, I.~Martel, M.~Milin, A.~Musumarra, M.~Papa, M.~G. Pellegriti,
  F.~Perez-Bernal, R.~Raabe, G.~Randisi, F.~Rizzo, G.~Scalia, O.~Tengblad,
  D.~Torresi, A.~Maira Vidal, D.~Voulot, F.~Wenander, and M.~Zadro.
\newblock Experimental study of the collision $^{11}${Be} + $^{64}${Zn} around
  the {C}oulomb barrier.
\newblock {\em Phys. Rev. C}, 85:054607, 2012.

\bibitem{CHB10}
P.~Capel, M.S. Hussein, and D.~Baye.
\newblock Influence of the halo upon angular distributions for elastic
  scattering and breakup.
\newblock {\em Phys. Lett. B}, 693:448, 2010.

\bibitem{HT03}
P.~G. Hansen and J.~A. Tostevin.
\newblock Direct reactions with exotic nuclei.
\newblock {\em Annu. Rev. Nucl. Part. Sci}, 53:219, 2003.

\bibitem{Gad08}
A.~Gade, P.~Adrich, D.~Bazin, M.~D. Bowen, B.~A. Brown, C.~M. Campbell, J.~M.
  Cook, T.~Glasmacher, P.~G. Hansen, K.~Hosier, S.~McDaniel, D.~McGlinchery,
  A.~Obertelli, K.~Siwek, L.~A. Riley, J.~A. Tostevin, and D.~Weisshaar.
\newblock Reduction of spectroscopic strength: Weakly-bound and strongly-bound
  single-particle states studied using one-nucleon knockout reactions.
\newblock {\em Phys. Rev. C}, 77:044306, 2008.

\bibitem{JHH11}
\O{}. Jensen, G.~Hagen, M.~Hjorth-Jensen, B.~Alex Brown, and A.~Gade.
\newblock Quenching of spectroscopic factors for proton removal in oxygen
  isotopes.
\newblock {\em Phys. Rev. Lett.}, 107:032501, 2011.

\bibitem{FOB12}
F.~Flavigny, A.~Obertelli, A.~Bonaccorso, G.~F. Grinyer, C.~Louchart,
  L.~Nalpas, and A.~Signoracci.
\newblock Nonsudden limits of heavy-ion induced knockout reactions.
\newblock {\em Phys. Rev. Lett.}, 108:252501, 2012.

\bibitem{YOM11}
M.~Yahiro, K.~Ogata, and K.~Minomo.
\newblock Eikonal reaction theory for neutron removal reaction.
\newblock {\em Prog. Theor. Phys.}, 126:167, 2011.

\bibitem{Ber05}
C.~A. Bertulani.
\newblock Relativistic continuum-continuum coupling in the dissociation of halo
  nuclei.
\newblock {\em Phys. Rev. Lett.}, 94:072701, 2005.

\bibitem{OB09}
K.~Ogata and C.~A. Bertulani.
\newblock Dissociation of relativistic projectiles with the
  continuum-discretized coupled-channels method.
\newblock {\em Prog. Theor. Phys.}, 121:1399, 2009.

\bibitem{OB10}
K.~Ogata and C.~A. Bertulani.
\newblock Dynamical relativistic effects in breakup processes of halo nuclei.
\newblock {\em Prog. Theor. Phys.}, 123:701, 2010.

\bibitem{CJN13}
P.~Capel, R.~C. Johnson, and F.~M. Nunes.
\newblock The ratio method: A new tool to study one-neutron halo nuclei.
\newblock {\em Phys. Rev. C}, 88:044602, 2013.

\end{thebibliography}

\end{document}